\documentclass[sigconf]{acmart}

\AtBeginDocument{%
  }

\setcopyright{rightsretained}
\copyrightyear{2025}
\acmYear{2025}
\acmDOI{TBD}

\acmConference[ICMI '25]{Proceedings of the 27th ACM International Conference on Multimodal Interaction}{October 13--17, 2025}{Canberra, Australia}
\acmBooktitle{Proceedings of the 27th ACM International Conference on Multimodal Interaction (ICMI '25), October 13--17, 2025, Canberra, Australia}
\acmISBN{TBD}

\graphicspath{ {./images/} }
\usepackage{enumitem}
\usepackage{float}

\begin{document}

\title{Attention, Action, and Memory: How Multi-modal Interfaces and Cognitive Load Alter Information Retention}

\author{Omar Elgohary}
\affiliation{%
\orcid{0009-0004-2778-5423}
  \institution{University of Minnesota}
  \city{Minneapolis}
  \country{USA}}
\email{elgoh003@umn.edu}

\author{Zhu-Tian Chen}
\affiliation{%
  \institution{University of Minnesota}
  \city{Minneapolis}
  \country{USA}}
\email{ztchen@umn.edu}

\begin{abstract}
Each year, multi-modal interaction continues to grow within both industry and academia. However, researchers have yet to fully explore the impact of multi-modal systems on learning and memory retention. This research investigates how combining gaze-based controls with gesture navigation affects information retention when compared to standard track-pad usage. A total of twelve participants read four textual articles through two different user interfaces which included a track-pad and a multi-modal interface that tracked eye movements and hand gestures for scrolling, zooming, and revealing content. Participants underwent two assessment sessions that measured their information retention immediately and after a twenty-four hour period along with the NASA-TLX workload evaluation and the System Usability Scale assessment. The initial analysis indicates that multi-modal interaction produces similar targeted information retention to traditional track-pad usage, but this neutral effect comes with higher cognitive workload demands and seems to deteriorate with long-term retention. The research results provide new knowledge about how multi-modal systems affect cognitive engagement while providing design recommendations for future educational and assistive technologies that require effective memory performance.

\end{abstract}

\ccsdesc[500]{Human-centered computing~User studies}

\keywords{Multi-modal interaction; Gaze tracking; Gesture recognition; Human-computer interaction (HCI); Learning retention; Cognitive load; Usability evaluation; Educational technology.}

\maketitle
\section{Introduction}
The mixture of eye-tracking and free-space gesturing as a means of interactivity has been researched rather thoroughly \cite{yuan2025survey}. Each modality has its advantages as well as disadvantages. For instance, the eye gaze easily and effortlessly reflects user focus and aids in the fast target acquisition, whereas free space gestures provide a variety dynamic change possibilities. In contrast, free-space gestures offer no feedback support and eye tracking suffers from the natural jitter and inaccuracy of fine control movements ~\cite{mirza2015gaze}.

Recent studies emphasize that integrating gaze and gesture systems produce smoother and finer interactions. This is due to the leveraging the strengths of each ~\cite{mirza2015gaze,wibirama2020spontaneous,yuan2025survey}. These results range from surpassing to matching the performance of the traditional techniques. However, these multi-modal systems introduce a higher cognitive load, as users must simultaneously manage multiple input channels, coordinate visual attention, and perform precise motor actions, often without explicit feedback mechanisms ~\cite{yuan2025survey}.

Cognitive Load Theory describes the constraints of working memory on a particular task and the stress it incurs while emphasizing the limited cognitive resources available ~\cite{yuan2025survey,kosch2023survery,kopp2017gesture}. It especially concerns itself with three types of load: intrinsic, which accompanies the complexity of the material to be learned, extraneous which is a result of the design and presentation of the materials as well as the structure of the activities, and germane load which is the effort exerted in recalling or learning the information~\cite{kopp2017gesture,sweller2011cognitive}. In turn, multi-modal interaction systems increase cognitive challenges with their complexity and require cross-modal coordination ~\cite{yuan2025survey,kosch2023survery}. Given the interest from the industry and academia, more research is needed on the impact of multi-modal interaction on cognitive processes such as learning and retention of information.

This paper seeks to assess whether the retention of users’ information is better with a multi-modal gaze and gesture interface compared to a standard track-pad interface. In order to assess this, participants were asked to engage with articles where important parts were textually shrunk, requiring greater usage of each system. Information recall metrics, cognitive load collected through the NASA-TLX, usability perception scores captured through the System Usability Scale, as well as behavioral metrics were collected as a baseline to analyze the effect multi-modal interactions have on cognitive performance.
The main contributions of this paper are:

\begin{itemize}[topsep=10pt]
\item Design and implement a multi-modal reading interface that integrates real-time gaze tracking and gesture-based control for article information acquisition.
\item Conduct a within-subjects user study comparing gaze and gesture interaction against traditional interfaces in terms of immediate and delayed information retention, cognitive load, and usability perceptions.
\item Present empirical evidence showing that multi-modal interaction is on-par with traditional interaction methods, even with increased mental demands.
\end{itemize}

\section{Related Work}
Related work to our gaze+gesture system and user study can be divided into two primary categories: multi-modal gaze+gesture input systems and cognitive load theory with its relation to information retention.

\subsection{Multi-modal Gaze and Gesture Systems}

There has been growing interest in leveraging multi-modal interactions that combine gaze and gesture input due to their potential to create more natural, expressive, and hands-free experiences. Recent advances in machine learning have enabled standard webcams to approximate the accuracy traditionally reserved for infrared (IR) eye-tracking systems, making gaze interactions increasingly accessible without expensive specialized hardware ~\cite{papa2016web,apple2024eyetracking,beamtracker2024,yuan2025survey}. At the same time, improvements in computer vision have made gesture and gaze recognition more intuitive and deployable with low-cost sensors, with these interfaces now appearing in domains such as the automotive industry, AR/VR headsets, and even general phone usage~\cite{kumar2022car,chen2022car,ken2017vr,apple2024eyetracking}.

Prior systems have explored gaze-supported or gesture-supported input individually, demonstrating benefits for target acquisition and remote control tasks ~\cite{mirza2015gaze}. Combined gaze and gesture interaction promises even greater expressiveness by using gaze to indicate attention and gestures to confirm or manipulate targets  ~\cite{mirza2015gaze}. However, most multi-modal implementations still run into issues, including gaze-tracking inaccuracies, the need for calibration, and gesture mis-recognition when deployed in unconstrained environments ~\cite{yuan2025survey}.

Although some systems have integrated gaze and gesture successfully for tasks such as large display interaction or remote collaboration, few works have considered how the cognitive implications of multi-modal input impacts learning and retention in users. Most prior studies focus on interaction performance without examining effects on user cognitive load, learning, and/or retention. In addition, many mainly rely on NASA-TLX assessments to draw their conclusions ~\cite{kosch2023survery}.

\subsection{Cognitive Load and Learning Retention in Multi-modal Systems}

Cognitive load considerations have become an integral component of planning any new multi-modal system. There have been numerous evaluation approaches, such as the NASA-TLX, that have been widely applied in the discipline. The NASA-TLX has tirelessly been recognized as the most important and widely accepted measure of user cognitive load in the industry. The more recent studies, such as the one conducted by Kosch et al. \cite{kosch2023survery}, were able to highlight quite effectively how NASA-TLX becomes an outdated model of HCI in multi-modal evaluation. In which Kosch et al. \cite{kosch2023survery}, argued that HCI multi-modal studies require more innovative and sophisticated evaluation methods that are not solely reliant on survey-based data. This is also the position taken by Longo et al.\cite{longo2022human} , who, having broadened the concept of mental workload, advocated for higher order synthesis of layered measurements toward precision in defining workload accuracy.

Developing newer techniques for evaluating cognitive workloads other than the NASA-TLX is gaining popularity. To cater to specific needs, tailored subjective approaches have been designed. Tracy et al. \cite{tracy2006measuring} are noted for integrating an audio-visual memory test with a tagging task and self-reported workload measure to evaluate workload pertaining in web-design. Likewise, Wilson et al. \cite{wilson2011surg} introduced the SURG-TLX for surgical contexts to capture task and situation complexity that transcends the bounds of NASA-TLX.

In an effort to examine the hypothesis that subjective evaluation of workload could be combined with objective measures of workload, studies showed that pupil dilation and heart-rate variability corresponds with NASA-TLX scores \cite{naik2022measurement,schuss2024heart} . Moreover, eye-tracking devices have been utilized to classify cognitive workload during tasks which presents instantaneous feedback regarding mental engagement with the activity\cite{kaczorowska2023eye}. 

To the best of our knowledge, we are some of the few to investigate retention quizzes as potential evaluative tools for cognitive load in multi-modal systems.

\section{The System}
To fully understand the system and its effects, here's a detailed breakdown of the components and interactions within the interface.

\begin{figure}[ht]
  \centering
  \includegraphics[
  width=0.5\textwidth,
  height=0.5\textheight,
  keepaspectratio
]{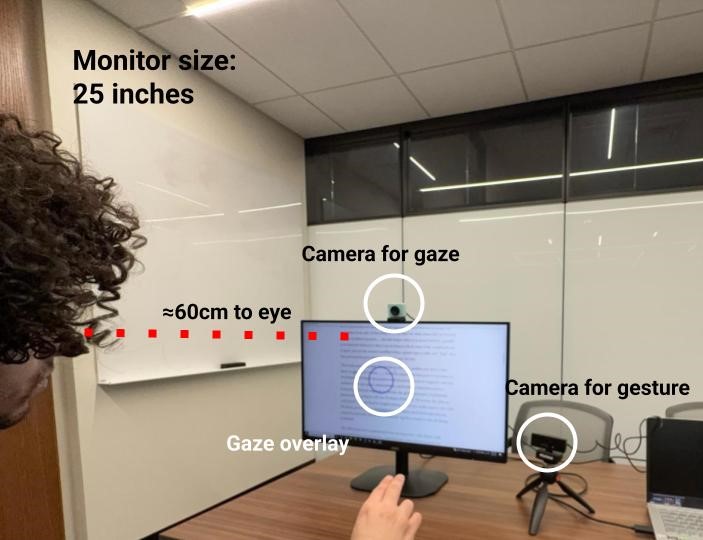}
  \caption{This is the apparatus used for our interaction
techniques and evaluation. The eye tracker is located
above the monitor, while the hand tracker is to the right/left of the user depending on handedness.}
  \label{fig:new_system}
\end{figure}

\subsection{Gaze Implementation}
For the gaze aspect of the interface we developed a custom gaze-tracking module based on the Beam Eye Tracker SDK ~\cite{beamtracker2024}, which is capable of real-time gaze estimation using low-cost webcams. Participants were seated about 60 cm from the screen and asked to optimally position themselves before the study session began. After initial calibration, participants were able to move their heads freely without, massive changes like standing up, for the duration of the task.  

The system provided a gaze overlay by displaying a continuous gaze "bubble" throughout the duration of the task. This provided relatively non-intrusive feedback without competition for cognitive resources. Gaze data was obtained as streaming data in real time using the Beam SDK TrackingListener class, where gaze coordinates were timestamped and stored in a CSV file for later analysis.

The reason we selected the Beam SDK implementation is because its tracking ability rivaled IR systems tracking while having accessible hardware requirements and lower cost than traditional infrared systems. This then allows us to better generalize our findings since most recent systems are primarily influenced by machine learning and/or most use some form of overlay to display detection. It is also important to point out that the stated margin of error is 1.5°~\cite{beamtracker2024}.

\subsection{Gesture Implementation}

\begin{figure}[ht]
  \centering
  \includegraphics[
  width=0.5\textwidth,
  height=0.5\textheight,
  keepaspectratio
]{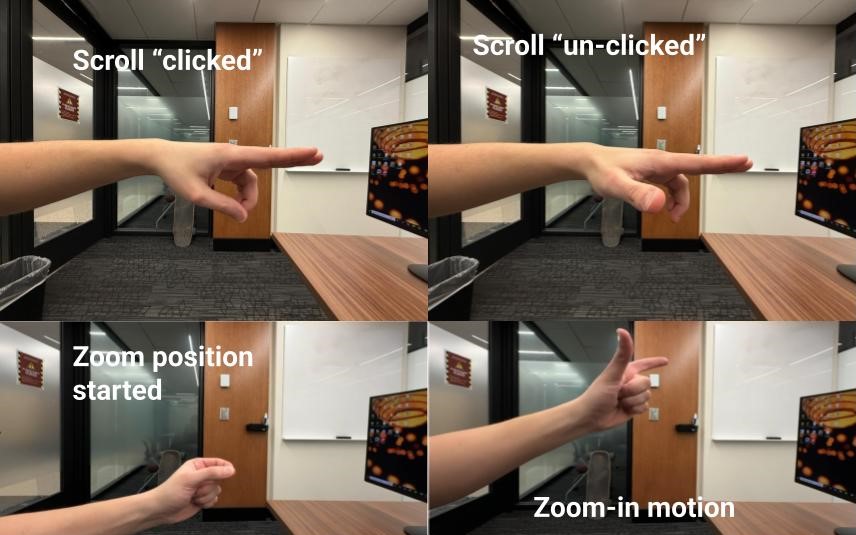}
  \caption{Example Motions}
  \label{fig:new_system}
\end{figure}

For detecting and classifying user gestures, we developed a custom gesture recognition module using OpenCV with MediaPipe Hands ~\cite{opencv_library}. The system allows two main types of gestures: one for scrolling and one for zooming in and out. Capture of the webcam input was done real-time at 30 FPS, and detection was done in real time.  

The scroll gesture was described as the sustained position of the index and middle finger of the participant extending in a horizontal fashion while pinching with thumb and ring finger together. This position simulates a drag and drop action whereby clicking was translated into vertical wrist movement which in turn scrolled up or down the reading material. The system detected the positions of finger landmarks, and whether the index finger and thumb were extended or held close together whereby the thumb and ring fingers were pinched with a small inter-finger distance. During the scroll gesture detection, the system calculated the change of wrist height (dy) to compute change in scroll amount. Early prototypes explored the implementation of scrolling through simple vertical waving motions of the hand. However, this approach proved to be inaccurate, as normal background hand movements during reading often were detected as a gesture that led to false positives. 

The zoom gesture implementation is a progressive movement. Users first pinched together their thumb and index fingers, signaling the beginning of a zoom action. The system interpreted the gesture as a zoom-in or zoom-out command based on the vertical motion with respect to holding the pinch. Specifically, zoom-in was triggered by upward movement while pinched followed by unpinching, and zoom-out was triggered by downward movement with unpinching. This finite-state approach ensured that zooming was only achieved when a pinch was detected, stabilized, and subsequently released with correct directional motion.  

An alternative simpler implementation using only a basic pinch gesture for zooming was also explored. This method suffered from poor accuracy, as the gestures for zooming in and zooming out were too similar in both position and motion trajectory. Zoom events depicting the nature of the gesture (scroll, pinch-in, and pinch-out), gesture location relative to the gaze position, and time stamp were saved on a CSV in real time for further behavioral evaluation.

Combined with gaze anchoring, this hands-free system enables participants to control reading content smoothly and intuitively in a mouse-free and keyboard-free environment.

\section{Study Setup}
\subsection{Participants}
Using mostly university posters, we managed to recruit 12 participants (P1–P12; M=8, F=4; Age: 18–28). Ten of the participants were right-handed while the other two were left-handed, so there was some variation in handedness. All participants were classified as having normal vision, or vision which could be corrected to normal. With the exception of P5, who claimed to own a Meta Quest3 device that has integrated gesture and gaze control mechanisms, all participants claimed to have no major prior experience interacting with systems using gaze and hand-based gestures. Additionally, P9 was the only participation who disclosed that they were diagnosed with ADHD.

Participants were compensated with a 10 dollar Amazon gift card for their time. Each participant completed both the gesture+gaze and track-pad interaction conditions twice, allowing within-subject comparisons.

\subsection{Procedure}
Each session commenced with a recap of what the motivation for the study was alongside the procedures, as well as a brief informed consent section. Then participants were presented with the gesture+gaze system and completed a structured training session of no more than 10 minutes, which included hands on training and a short trial run to ensure system operability. There were four articles that participants read as part of the study. For each article, participants either interacted with the reading interface using the gesture+gaze system or a standard laptop track-pad. The system condition, designated as either gesture+gaze or track-pad, and the article allocated were both randomized across sessions in order to mitigate potential ordering biases as well as difficulties regarding article complexity.  

In advance of every session that involved the gesture+gaze system as the reading device, participants completed a session calibration procedure to maintain adequate gaze tracking accuracy. For every reading trial, participants were allotted a maximum of 8 minutes to complete reading an article that was rated to take around four minutes to read~\cite{gelitz2025soccer,fischetti2024pickleball,isbister2017fidget,guglielmi2025bmi}. Participants could advance at any point if they felt they had finished prior to the set time. After every article, participants were asked to complete a short retention quiz made up of 3 multiple choice, 3 true/false, and 3 short answer questions. This cycle of reading and quiz was conducted four times with each participant.

After finishing all reading and quizzing activities, respondents were registered for completing a System Usability Scale questionnaire, a NASA-TLX cognitive workload index, and a short semi-structured questionnaire about their feedback. To measure information retention over longer periods, respondents were invited the subsequent day to undertake another set of four quizzes associated with the four articles they were assigned to in the previous session.

\section{Results}
\subsection{Retention Performance}
\begin{figure}[h]
  \centering
  \includegraphics[width=\linewidth]{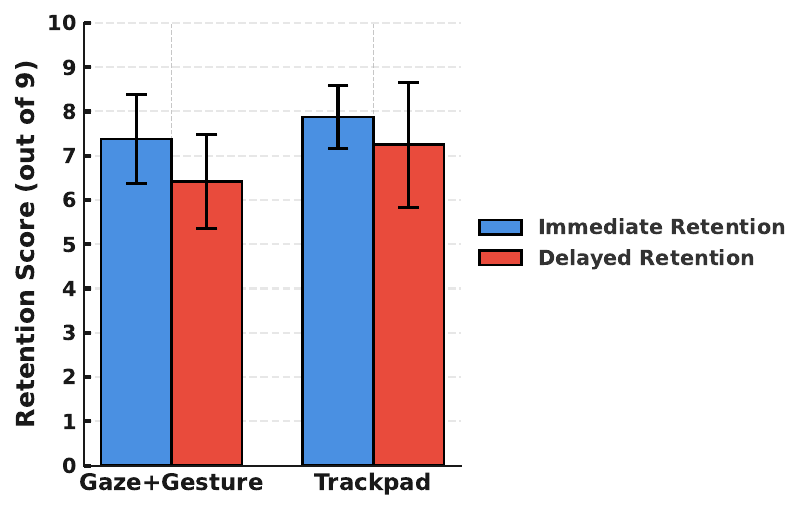}
  \caption{Immediate and delayed retention scores for gaze+gesture and Track-pad systems. Error bars represent standard deviations.}
  \label{fig:ret}
\end{figure}

Figure~\ref{fig:ret} displays the mean immediate and delayed retention scores for gaze+gesture and Track-pad systems. Participants seemed to score slightly higher in immediate retention with the track-pad system (M = 7.88, SD = 0.71) as opposed to gaze+gesture (M = 7.38, SD = 1.00). However, a paired-samples t-test indicated that the difference was not significant, $t(11) = -2.03$, $p = 0.067$, although some tendency provided evidence of a trend.

On the contrary, the delayed retention scores were significantly higher with the track-pad (M = 7.25, SD = 1.41) than with gaze+gesture (M = 6.42, SD = 1.06). This difference has been confirmed as statistically significant by a paired-samples t-test, $t(11) = -2.87$, $p = 0.015$, and also sustained by a Wilcoxon signed-rank test, $W = 0.0$, $p = 0.011$.

These results imply that while the systems used did not differ in immediate information retention, the difference was notable in long-term retention where participants interacted with the articles through the track-pad interface.

Participants required roughly the same reading time with both systems, averaging 277.98 seconds (about 4.63 minutes) per article for gaze+gesture and 281.04 seconds (about 4.68 minutes) for track-pad. This minimal difference in reading time indicates that the retention effects noted were not just the outcome of the different time-on-task, but more likely the result of the differences in the interaction demand of the modality itself.

\subsection{Cognitive Load and Usability Results}

\begin{figure}[h]
  \centering
  \includegraphics[width=\linewidth]{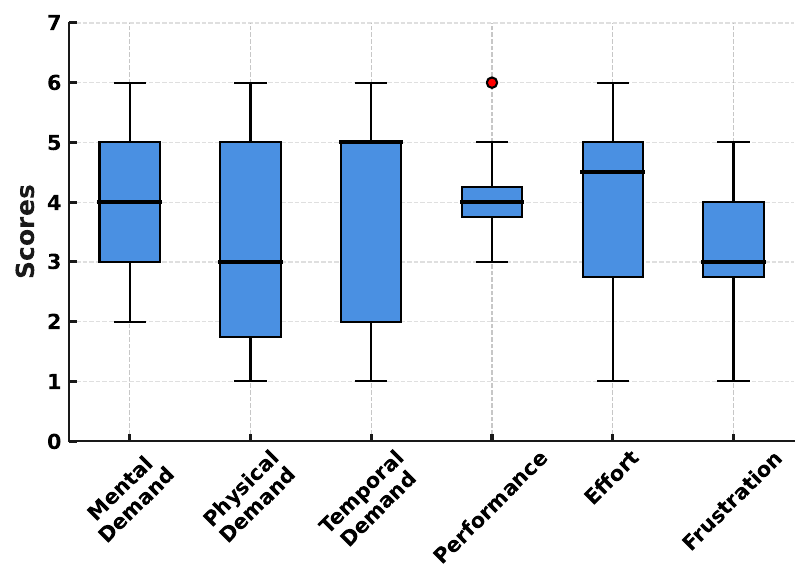}
  \caption{Distribution of NASA-TLX subscale scores for the Gaze+Gesture system.}
  \label{fig:nasa_tlx_boxplot}
\end{figure}

Participants' perceived cognitive workload was measured with the NASA-TLX questionnaire in its six sub-scale divisions. Participants provided the following mean scores: Mental Demand (M = 4.00, SD = 1.21), Physical Demand (M = 3.25, SD = 1.91), Temporal Demand (M = 3.92, SD = 1.88), Performance (M = 4.08, SD = 0.90), Effort (M = 3.92, SD = 1.56), and Frustration (M = 3.17, SD = 1.11). These results show that the participants interacting with the gaze+gesture system, on average, reported a moderate workload from their interactions with the system, with relatively lower ratings for physical effort and frustration compared to mental demands.  

System usability was assessed with the SUS questionnaire. The average SUS score for the gaze+gesture interface was 62.29 (SD = 14.87) out of 100. This indicates a moderate perception of usability and falls well under the often-cited benchmark of 68 which is usually viewed as the cutoff for good usability \cite{bangor2009sus}. In summary, participants perceived the system to be functioning but not particularly straightforward or effortless for use.

\subsection{Gesture and Gaze Behavior Analysis}

\begin{figure}[h]
  \centering
\includegraphics[width=\linewidth]{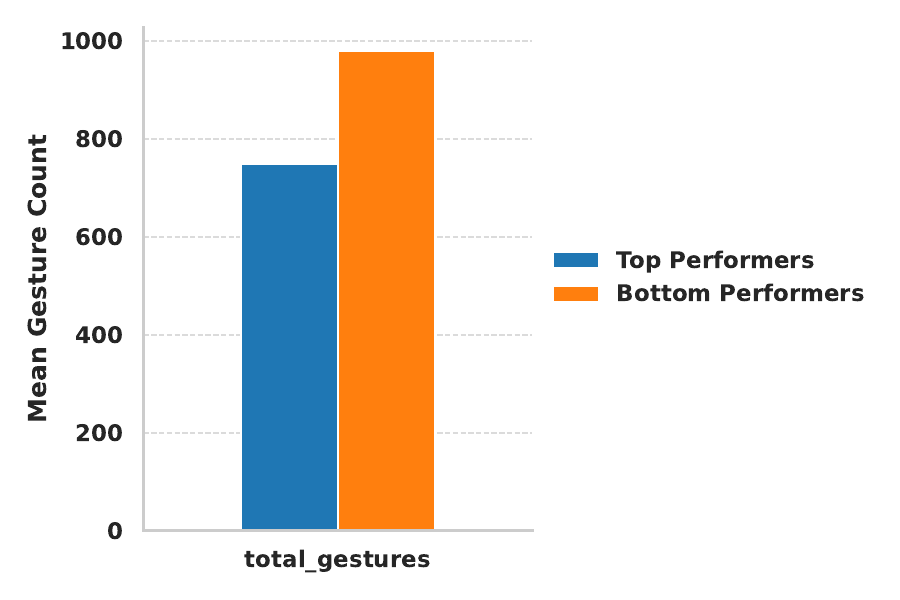}
  \caption{\textbf{Mean gesture counts for top and bottom retention performers.}} 

  \label{fig:top_vs_bottom}
\end{figure}

Quantitative analysis of gesture use revealed a particularly sharp gap between top and bottom retention performers. Participants with lower retention scores made a total number of gestures that was much higher than their higher performing counterparts. Bottom performers, on average, made roughly 30\% more gestures than top performers which insinuates that the intensity of interaction was counterproductive in this instance. Instead, overdependence on gestures, scrolling for instance, may have indicated greater task difficulty or lower reading efficiency, which in turn, would increase cognitive load without improving the retention of information. 

Apart from gesture usage, metrics related to gaze tracking also shed more light regarding the participants’ levels of attention. Most participants in all sessions completed their interactions with the system displaying low gaze variance paired with even lower average gaze movements which depict concentrated and sustained attention as well as monotonous task engagement. A small subset of participants, however, displayed low averages of extreme high variance and large changes in gaze between gestures suggesting possible episodes of off-task attention or difficulty keeping a visual focus on areas that are relevant to the work at hand.

All of the above findings imply that outcomes associated with retention were poorer in the case of higher gestural interaction and unstable visual attention. Through the reading tasks, participants who were less active with gestures and fixated their gaze more tended to recall more information. This reinforces the overarching assumption that learning through the interaction of multiple modalities does not rely solely on the richness of inputs such as gestures and gaze; rather, it requires the structured and purposeful control of those inputs.

\subsection{Open-ended Short Interviews }

Following the tasks, the participants went through a set of short interviews which were open-ended in nature, aiming to capture how they interacted with the gaze+gesture system in comparison to the track-pad setup. Their answers were aggregated into overarching themes which we describe below.

\subsubsection*{Retention and Understanding}
Most participants claimed that the use of the track-pad aided their comprehension and recall from memory significantly. A lot of them considered the gaze+gesture system to be very attention capturing or distracting. Many claimed that the gaze overlay in addition to the need to perform gestures made reading less simple and therefore wasn’t easy to read on. Problems with backward navigation, as well as discomfort with the zoom gesture, were also frequently noted. Only a few participants credited some benefits to the gaze-based interface, often attributing this to more focus being exerted on some specific words, though even these participants did admit that there were additional cognitive difficulties.

\subsubsection*{Mental Effort}
All of the participants selected that the gaze+gesture control system required more mental effort than traditional controls. The need to think about one's palm and execute gestures as it was configured added to the mental effort. Specifically, scrolling gestures appeared to require a lot of mental effort and zoom gestures were usually awkward or difficult to activate with certainty. Some users with previously basic experience of using gestures on machines (like VR users) were able to adjust very quickly, but for most, the system was always a few steps ahead of the reading task.  

\subsubsection*{Naturalness}
The range of perceptions with regards to naturalness also diverged quite a bit. The idea of using hand motion for scrolling was accepted as intuitive by a majority of the participants, however, the overlay gaze itself was deemed distracting as well as unnatural. Still, one participant shared that he felt the overlay was beneficial for his ADHD. A few participants stated that they could improve their perception of naturalness if they were provided with longer practice sessions.  

\subsubsection*{Suggestions for Improvement}
Use of the gaze+gesture has offered countless improvement ideas which  all generally fit into a theme. Most of the ideas lean towards decreasing the size and gaze overlay visibility. Other suggestions include refining the methods through which zooming is recognized, allowing the user to set the amount of gaze feedback, and granting the ability to change the gestural incremental levels of scrolling and zooming.

\subsubsection*{Usage Contexts}
Participants noted a few distinct contexts where a gaze+gesture interface would be especially helpful. Examples comprised of accessibility solutions for the physically challenged, educational settings meant to improve learner engagement, media viewing from unconventional positions (such as TV screen without a desk), and professional environments with video consultations where free use of hands is needed, like in work 3D modeling, video editing, and working with multiple monitors. Some participants are of the opinion that while the system may not aid in conventional controls for more tasking works, it does provide useful value in more specialized case scenarios.

\section{Discussions}

Our research indicates that although multi-modal gaze+gesture interaction achieves information retention on par with the track-pad interface's immediate recall, it adds an extra layer of cognitive workload, increasing usability issues that impact long-term information retention. Most users stated that the dual gaze and gesture input paradigm was more mentally taxing, which, at times, pulled focus away from the main reading activities. These remarks reflect previous research focusing on the cognitive constraints within multi-modal systems \cite{yeji2024impact,kosch2023survery}.  

Surprisingly, retention testing that involved immediate and delayed quizzes turned out to be a valid stand-in for estimating cognitive load. Participants who described greater difficulty concentrating or using the gaze+gesture system tended to have sharper declines in retention over time, demonstrating that information recall can serve as a reflection of cognitive demands. While cognitive load is usually evaluated with subjective tools such as NASA-TLX, our findings restore the hypothesis that retention decay, a more objective measure, can be used alongside subjective metrics to provide a fuller representation of user experience within multi-modal systems.

Both the quantitative metrics and qualitative assessments revealed an important user experience insight: users with better retention tended to use fewer gestures and exhibited more stable gaze. As such, less is often more when it comes to uncluttered, coherent structure, while the overall system interaction discipline is key for keeping cognitive focus under control during learning tasks. Educational applications for future designs of multi-modal interfaces might be more effectively served through non-promotive, directed interactions rather than unlimited interface control.

In addition, while most participants reported that scrolling gestures became intuitive after some training, the gaze overlay received wide disapproval as being too distracting. Among participants, a prominent suggestion was to make the gaze visualization an optional feature or less pronounced, emphasizing the urgency of addressing feedback channel designs in gaze-augmented interfaces.

\subsection{Limitations}
A few limitations from this research should be mentioned. To start, the small sample size (N=12) impacts the scope of generalizability of our findings. Studies with different sample compositions will be important to confirm these findings among other populations. While we provided some training on the gaze+gesture system, participants likely had far more lifetime experience using a track-pad, which could have skewed results toward the traditional interface. Users may surpass the initial learning curve with multi-modal systems and reap benefits after enduring extended periods of training, which longitudinal studies are likely to capture. 

Lastly, this research examined article reading, but other types of tasks such as problem-solving or creative tasks could exhibit different patterns and different degrees of effectiveness with multi-modal interactions.

\subsection{Future Work}
Based on our findings, there are several additional promising avenues to pursue. One focuses on the optimization of gaze visualization techniques, testing the use of subtler pointers or adaptive overlays that portray gaze but do not distract. Another equally important focus is the enhancement of gesture sets to streamline complex actions, for instance, replacing current pinch-based zooming gestures with more discrete, effortless gestures. 

In addition, systems could provide multi-level feedback capabilities that adjust to the user's mental effort in real-time, enabling fewer steps to interact with the system when mental exertion is at its highest. Finally, the study of multi-modal systems would benefit from more advanced levels of learning beyond memorization or detail work, including abstract reasoning, to help define where supportive gaze and gesture interfaces would be most helpful for users.

\section{Conclusion}
This research explored the impact of multi-modal gaze+gesture interaction relative to a traditional track-pad interface in learning, usability, and cognitive load during reading tasks. Although the immediate recall scores of the two interfaces were roughly comparable, qualitative and behavioral metrics indicated critical differences beneath the surface. Users of the gaze+gesture system reported heightened cognitive effort, greater focus maintenance difficulties, higher gesture and gaze "idling", all of which lowered the delayed retention performance. As such, the results indicate that while short-term outcomes may be achieved with multi-modal systems, the systems could impose additional cognitive burdens that hamper deep learning.  

As noted, a focus of the research study was to examine the implications of few objective behavioral metrics as contrasted to subjective measures like NASA-TLX. Usability metrics such as gaze variation and retention decay were found to extend traditional cognitive load boundaries, proving useful. Retention testing, especially, proved to be effective measuring cognitive strain in interactive system studies.  

Participant discourse did highlight some advantages for gaze + gesture systems in more niche applications, which included: specialized accessibility use case scenarios, educational environments, and hands-free media consumption. Moreover, participants put forth commendable design recommendations, notably segmenting gaze overlays, gesture simplification, cognitive load responsiveness, and more adaptive interfaces.

Broadly speaking, these results highlight the importance of good multi-modal interaction design. Further research should work towards achieving the right balance between expressivity and cognitive demand, making sure that multi-modal systems do not impede users’ ability to concentrate, learn, and engage meaningfully with content. Moreover, this study shows the extent to which retention quizzes can practically supplement more subjective cognitive load assessments that rely on traditional methodologies. Actions like immediate and delayed recall as well as self-reporting provide richer understanding of the cognitive workload interactive systems impose, making it vital that self-report methodologies incorporate these alongside the usual metric for user experience evaluation.

\section*{Safe and Responsible Innovation Statement}
This study analyzes the use of gaze and gesture based reading interfaces and their effects on a user’s state of cognition. As gaze data can sensitive information pertaining to a user’s cognitive state, any implementation must ensure absolute data privacy as well as clear communication about data collection and utilization policies. This research has found that multi-modal systems can deliberately enhance the cognitive load of the user, indicating that exaggerated design is not recommended in educational systems and in the context of user accessibility. We emphasize the need for inclusive design of multi-modal systems where interface adaptation is catered to each user. By emphasizing ethical use and minimizing cognitive strain, we aim to guide responsible deployment of multi-modal systems in educational and assistive contexts.

\bibliographystyle{ACM-Reference-Format}
\bibliography{citations}
\newpage
\onecolumn 

\section*{Appendix}

\subsection*{Retention Scores}

\begin{table}[H]
  \centering
  \caption{Per-participant immediate and delayed retention scores (raw scores out of 9 with corresponding percentages) for Gaze+Gesture (G+G) and Track-pad systems.}
  \label{tab:participant_retention_scores}
  \begin{tabular}{lcccc}
    \toprule
    P & G+G & Track-pad & Delayed G+G & Delayed Track-pad \\
    \midrule
    P1 & 9.0 (100.0\%) & 8.5 (94.4\%) & 7.5 (83.3\%) & 8.0 (88.9\%) \\
    P2 & 6.0 (66.7\%) & 8.0 (88.9\%) & 6.0 (66.7\%) & 7.0 (77.8\%) \\
    P3 & 8.5 (94.4\%) & 8.5 (94.4\%) & 7.5 (83.3\%) & 7.5 (83.3\%) \\
    P4 & 7.5 (83.3\%) & 7.5 (83.3\%) & 7.0 (77.8\%) & 7.5 (83.3\%) \\
    P5 & 8.5 (94.4\%) & 8.5 (94.4\%) & 8.0 (88.9\%) & 8.5 (94.4\%) \\
    P6 & 7.0 (77.8\%) & 8.5 (94.4\%) & 7.5 (83.3\%) & 9.0 (100.0\%) \\
    P7 & 7.5 (83.3\%) & 8.0 (88.9\%) & 5.5 (61.1\%) & 9.0 (100.0\%) \\
    P8 & 8.0 (88.9\%) & 8.0 (88.9\%) & 6.0 (66.7\%) & 7.5 (83.3\%) \\
    P9 & 6.5 (72.2\%) & 6.5 (72.2\%) & 6.5 (72.2\%) & 7.5 (83.3\%) \\
    P10 & 6.5 (72.2\%) & 6.5 (72.2\%) & 5.0 (55.6\%) & 5.0 (55.6\%) \\
    P11 & 6.0 (66.7\%) & 8.0 (88.9\%) & 5.0 (55.6\%) & 5.0 (55.6\%) \\
    P12 & 7.5 (83.3\%) & 8.0 (88.9\%) & 5.5 (61.1\%) & 5.5 (61.1\%) \\
    \bottomrule
  \end{tabular}
\end{table}

\subsection*{Reading Times}

\begin{table}[H]
  \centering
  \caption{Average reading time (in minutes) per participant for Gaze+Gesture and Track-pad systems.}
  \label{tab:participant_times}
  \begin{tabular}{lcc}
    \toprule
    Participant & Avg Time G+G (min) & Avg Time Track-pad (min) \\
    \midrule
    P1 & 5.56 & 4.98 \\
    P2 & 4.01 & 2.68 \\
    P3 & 5.41 & 4.77 \\
    P4 & 6.65 & 5.88 \\
    P5 & 3.94 & 4.83 \\
    P6 & 4.97 & 5.08 \\
    P7 & 2.72 & 2.58 \\
    P8 & 5.93 & 6.09 \\
    P9 & 5.50 & 8.00 \\
    P10 & 2.31 & 1.85 \\
    P11 & 3.61 & 5.09 \\
    P12 & 4.89 & 4.38 \\
    \bottomrule
  \end{tabular}
\end{table}

\end{document}